\documentclass[prb,showpacs,twocolumn,superscriptaddress]{revtex4}
\usepackage{amsmath,graphicx,latexsym}

\begin{document}

\title{Accurate polarization within a unified Wannier function
formalism}

\author{Massimiliano Stengel}
\author{Nicola A. Spaldin}

\affiliation{Materials Department, University of California, Santa Barbara, 
             CA 93106-5050, USA}

\date{\today}

\begin{abstract}
We present an alternative formalism for calculating the 
maximally localized Wannier functions in crystalline solids, obtaining an
expression which is extremely simple and general. 
In particular, our scheme is exactly invariant under
Brillouin zone folding, and therefore it extends trivially to the 
$\Gamma$-point case.
We study the convergence properties of the Wannier functions, their quadratic
spread and centers as obtained by our simplified technique.
We show how this convergence can be drastically improved by a simple 
and inexpensive ``refinement'' step, which allows for very efficient and
accurate calculations of the polarization in zero external field.
\end{abstract}	     


\maketitle


\section{Introduction}

The representation of the one-particle electronic structure of molecules
and solids in terms of localized Wannier~\cite{wannier} orbitals is 
nowadays ``enjoying a revival''~\cite{rmmartin} as a useful tool for many 
applications~\cite{mv,hydrogen,sharma,cangiani,
fernandez,posternak,dovesi,pavarini,fabris,feliciano,danish}.
The main impetus for this renewal of interest was given by the establishment,
by King-Smith and Vanderbilt (KSV),
of a formally exact relationship between the sum of the Wannier function (WF)
centers and a gauge-invariant Berry phase, in the context of the modern theory 
of polarization~\cite{kingsmith}. 
However, the intrinsic nonuniqueness in the Wannier function definition, and
the difficulty in defining their centers within a periodic cell calculation,
limited their practical use, until a particularly elegant method due to
Marzari and Vanderbilt~\cite{mv} (MV) became available some years ago.
Their scheme allows one to obtain, in a given isolated or extended system,  
a unique set of maximally localized Wannier functions that 
minimizes a well-defined spread functional. 
Moreover, the MV formalism provides as an important byproduct the positions of 
the WF centers, whose sum gives direct access to the macroscopic polarization
of the physical system.
%
%
The MV scheme, which became instantly popular, presents nevertheless an 
inconvenience, in that crystalline solids are treated on a different footing
with respect to the case of, e.g., large disordered systems simulated
at the $\Gamma$ point only.
The two prescriptions are indeed equivalent in the thermodynamic limit,
but they formally differ when discrete Brillouin zone (BZ) samplings 
(or finite supercells for isolated objects) are used, 
which is necessarily the case in any practical calculation.
In the first part of this work we show that there is nothing fundamental in 
this discrimination, i.e. that a given choice for a spread functional in
$\Gamma$-sampled cells dictates unambiguously the mathematical expression
in \emph{discrete} $k$-point space, and that invariance under 
``BZ folding'' is the guideline which estabilishes the link.
The resulting formalism is completely general and, while being
similar in spirit to the original (MV) one, presents a much simpler algebra.

A more relevant issue affects more directly the modern theory of polarization,
and concerns the asymptotic convergence with respect to BZ sampling.
It was already shown formally and numerically~\cite{mv,puma} that
both methods (KSV and MV) for calculating the polarization of a molecule or
a crystal in periodic boundary conditions (PBC) are plagued by a 
slow $\mathcal{O}(L^{-2})$ convergence, where $L$ is the linear dimension of the 
supercell containing the isolated molecule, or alternatively the resolution
of the $k$-point mesh.
The problem is usually addressed, in the context of the Berry-phase KSV
approach, by refining the $k$-point grid along ``stripes'' in the Brillouin 
zone within a separate, non-selfconsistent calculation~\cite{kingsmith}.
For finite systems, an extrapolation technique
was recently proposed~\cite{sagui}, in which the $\mathcal{O}(L^{-2})$ error is
removed from the Wannier multipoles by performing a series of calculations
in cubic supercells of 
increasing size~\footnote{A similar extrapolation technique was also proposed
  in the context of the discrete Berry-phase method~\cite{davidextr}}.
Both solutions are somewhat unsatisfactory, in that they require many
calculations to be performed on the same system, with a cost that is
higher than what is normally needed to converge total energies and
densities.
In the second part of this work we propose a simple 
``refining'' procedure, which is able 
to provide an extremely accurate value for the center and spread once
a well-localized set of maximally localized Wannier functions is available.
We show first formally and then by numerical examples 
that this technique, while requiring a very minor computational 
effort, is able to outperform in terms of accuracy both the standard 
KSV Berry-phase approach and the alternative formula based on 
``unrefined'' WF centers~\footnote{
  Another group, working in a different WF formalism~\cite{dovesiwannier},
  already noticed~\cite{dovesiknbo3} incidentally that the WF-based expression 
  for polarization can potentially provide better numerical convergence 
  than the Berry-phase approach.}.
Finally, our derivation also provides a novel, intuitive interpretation of the 
position/localization operator in periodic boundary conditions and of its 
relationship to the corresponding well-known free-space operators.


\section{Method}

The theoretical basis for the MV approach rests on a continuum formulation,
in which the space is infinitely extended in all directions; this translates
to an infinitely dense Brillouin zone sampling in the case of crystalline
solids.
For practical calculations a finite sampling (or finite simulation supercell)
is necessary, and MV give detailed prescriptions for the ``discretization'' of
the relevant mathematical objects (gradients and Laplacians in $k$-space).
We start here our alternative derivation from a slightly different 
viewpoint, i.e. we ``discretize'' the problem from the very beginning by 
choosing an appropriate spread functional in the $\Gamma$-point case,
and then work out the formulas in $k$-space without making any further
approximation.
This approach leads automatically to a general and size-consistent formalism,
that is invariant under BZ unfolding.

We assume a Born-von K\'arm\'an (BvK) supercell of 
volume $V_{BvK}$, which is a multiple of the primitive (P) unit cell of
the crystal (of volume $V_P$) under study.
In this system with periodic boundary conditions (PBC) there are $N$ allowed
Bloch vectors, where $N$ is given by the ratio between the volumes:
$$
N = \frac{V_{BvK}}{V_P}.
$$
The generalized Bloch orbitals (which are not necessarily
eigenstates of the Hamiltonian) are orthonormal on the primitive cell:
$$
\int_P \psi_{m\mathbf{k}}^*(\mathbf{r})
       \psi_{n\mathbf{k}} (\mathbf{r}) d \mathbf{r} = \delta_{mn},
$$
and can be written as usual:
$$
\psi_{n\mathbf{k}} (\mathbf{r}) = e^{i \mathbf{k}. \mathbf{r}}
u_{n \mathbf{k}} (\mathbf{r}),
$$
where the $u_{n \mathbf{k}}$ are periodic functions, and can be represented
on the reciprocal lattice of the P cell:
$$
u_{n \mathbf{k}} (\mathbf{r}) = \frac{1}{\sqrt{V_P}} \sum_{|\mathbf{G+k}|^2<
E_{cut}} e^{i\mathbf{G.r}} \tilde{u}_{n \mathbf{k}} (\mathbf{G}).
$$
$E_{cut}$ represents the plane-wave cutoff, while $\tilde{u}_{n \mathbf{k}}
(\mathbf{G})$ is the Fourier coefficient 
of the lattice-periodic part of the Bloch function:
$$
\tilde{u}_{n \mathbf{k}} (\mathbf{G}) = 
\frac{1}{\sqrt{V_P}} \int_P e^{-i\mathbf{G.r}}
 u_{n \mathbf{k}} (\mathbf{r}) d\mathbf{r}.
$$ 
We will use the BvK supercell for representing our
Wannier functions: 
$$
w_n(\mathbf{r}) =\frac{1}{N}
                 \sum_\mathbf{k} \psi_{n\mathbf{k}} (\mathbf{r}),
$$
where the normalization constant is chosen so that these $w_n$ are orthonormal on the
BvK supercell. A particularly simple relationship holds in
reciprocal space:
\begin{equation}
\tilde{w}_{n}(\mathbf{G+k}) = \frac{1}{\sqrt{N}}
                              \tilde{u}_{n\mathbf{k}}(\mathbf{G}).
\label{wfgspace}
\end{equation}
We remind the reader that the reciprocal lattice of the BvK cell is
spanned by all vectors of type $\mathbf{b}=\mathbf{G+k}$, which we will
call $\mathbf{b}$ in the following, to distinguish them from the $\mathbf{G}$
vectors of the primitive reciprocal lattice.

Eq.~\ref{wfgspace} does not define a unique set of Wannier functions, because 
of the gauge arbitrariness in the choice of the unitary representation of
the Bloch vectors.
This indeterminacy can be solved by defining a spread functional $\Omega$ which
depends explicitly on the gauge, so that the minimization of $\Omega$ leads to
a well defined set of localized orbitals with the desired properties.
Berghold and coworkers~\cite{berghold} proposed a particularly simple and 
appealing expression for $\Omega$ and the related Wannier 
centers $\bar{\mathbf{r}}_n$, which is valid for $\Gamma$-only BZ sampling
in a lattice of general symmetry.
Since our BvK supercell is sampled at $\Gamma$ by construction,
we can use the same expressions as Berghold, that in our notations read:
\begin{subequations}
\begin{eqnarray}
\label{eqwancen}
\mathbf{\bar{r}_n} & = & \sum_i \bar{\mathrm{w}}_i \mathbf{b}_i \mathrm{Im} 
\ln z_n^{(i)} \\
\Omega  &=& \sum_n \sum_i \bar{\mathrm{w}}_i \, 2 \, 
\big(1 - | z_n^{(i)}| \big).
\label{eqspread}
\end{eqnarray}
\end{subequations}
Here $z_n^{(i)}$ are dimensionless complex numbers given by:
$$
z_n^{(i)} = \langle w_n | e^{i \mathbf{b}_i .\mathbf{r}} |w_n \rangle = 
|z_n^{(i)}| e^{i\phi_n^{(i)}},
$$
and \{$\mathbf{b}_i,\bar{\mathrm{w}}_i$\} represents a small set of reciprocal lattice 
vectors $\mathbf{b}_i$ with weights $\bar{\mathrm{w}}_i$.
In the case of a cubic BvK supercell of edge $L$ these quantities
reduce to the $i=1,...3$ primitive reciprocal-space vectors of the BvK
cell and the weights are all equal:
$$
\mathbf{b}_i = \frac{2\pi}{L}\mathbf{\hat{i}}, \qquad 
\bar{\mathrm{w}}_i = \Big(\frac{L}{2\pi}\Big)^2,
$$
while in the most general case of a triclinic cell a maximum 
number of six independent vectors and weights must be used, according
to the prescriptions given in Ref.~\onlinecite{mv} and
~\onlinecite{silvestrelli}.

With these notations and conventions in hand, we are now ready to write
down a $k$-space expression for $\mathbf{\bar{r}}_n$ and
$\Omega$. Both quantities depend directly on $z_n^{(i)}$, and the key
of the derivation is then the ``Brillouin-zone unfolding'' of 
this latter quantity.
Using the same notation as MV:
$$
M^{(\mathbf{k,b_i})}_{mn} = \langle u_{m\mathbf{k}} |
                                    u_{n\mathbf{k+b}_i} \rangle, 
$$
it is straightforward to derive a very simple expression for $z_n^{(i)}$:
$$
z_n^{(i)} = \frac{1}{N} \sum_\mathbf{k} M^{(\mathbf{k,-b}_i)}_{nn}.
$$
With this formula we can write our operational definitions of position
and quadratic spread in $k$-space:
\begin{subequations}
\label{eqnsk}
\begin{eqnarray}
\label{wancenk}
\mathbf{\bar{r}_n} & = & -\sum_i \bar{\mathrm{w}}_i \mathbf{b}_i \mathrm{Im} 
\ln \Big[
\frac{1}{N} \sum_\mathbf{k} M^{(\mathbf{k,b}_i)}_{nn} \Big], \\
\Omega  &=& \sum_n \sum_i \bar{\mathrm{w}}_i \, 2 \, \big(1 - \Big| 
          \frac{1}{N} \sum_\mathbf{k} M^{(\mathbf{k,b}_i)}_{nn}
          \Big| \big)
\label{spreadk}
\end{eqnarray}
\end{subequations} 

It is interesting to notice the strikingly close similarity between our 
expression for the centers (Eq.~\ref{wancenk}) and Eq. 31 of MV,
the only difference being the order in which the complex logarithm and
the average over $k$-points is taken~\footnote{
  Our expression for the spread (Eq.~\ref{spreadk}) is also very close to Eq. 23 of
  MV (which is further discussed in~\onlinecite{posternak}), and coincides with it 
  when $\mathbf{\bar{r}_n}=\mathbf{0}$.
  }.
We argue that the one proposed here is a more natural choice, since 
it retains the correct translational properties of their formula,
while strictly enforcing size consistency.
Size consistency means that the formalism gives mathematically identical 
answers for the $k$-point representation and for the equivalent BvK 
real-space $\Gamma$-point representation. Our formula is correct by 
construction, and extends exactly to the case of isolated systems with 
$\Gamma$-point sampling without any further algebra.

Another advantage of our scheme is its simplicity, which becomes evident 
when taking the gradient of the spread functional with respect to an
infinitesimal unitary rotation in a given $\mathbf{k}$ subspace.
Thus, we consider the transformation:
$$
u_{n \mathbf{k}}' (\mathbf{r}) = \sum_{m} 
u_{m \mathbf{k}} (\mathbf{r}) U^{(\mathbf{k})}_{mn}, 
$$
where the rotation matrices $U^{(\mathbf{k})}$ are obtained by adding an
infinitesimal antiHermitian matrix $dW$ to the identity:
$$
U^{(\mathbf{k})} \sim 1 + dW^{(\mathbf{k})}.
$$
The variation of the total spread with respect to this transformation
is readily obtained in terms of the $M^{(\mathbf{k,b}_i)}$ matrices
and the \emph{phases} of the $z_n^{(i)}$ complex numbers:
$$
\Big( \frac{d \Omega}{dW^{(\mathbf{k})}} \Big)_{mn} = \frac{1}{N}
\sum_i \bar{\mathrm{w}}_i \Big( M^{(\mathbf{k,b}_i)}_{mn} C_n^{(i)*} +
$$
\begin{equation}
           + M^{(\mathbf{k-b}_i,\mathbf{b}_i)*}_{nm} C_n^{(i)} 
                 \Big) - H.c., 		       
\label{eqgrad}
\end{equation}
where $H.c.$ stays for Hermitian-conjugate, and $C_n^{(i)}$ is the
phase:
$$C_n^{(i)} = e^{i\phi_n^{(i)}} = \frac{z_n^{(i)}}{|z_n^{(i)}|}.$$
This expression for the gradient can easily be obtained by observing 
that, in Eq.~\ref{eqspread}, one can write $|z| = z e^{-i\phi}$.

%
We note that the for the spread functional (Eq.~\ref{eqspread}) several
possibilities exist, which are all equivalent in the thermodynamic 
limit~\cite{berghold}.
In the Appendix we briefly consider these alternatives, and we provide
a formal derivation of the gauge-invariant part of the spread~\cite{mv},
which further evidences the close relationship of our formulation to the 
original MV scheme.
Because of the exact mapping between the BvK supercell and the 
primitive one, we find it particularly natural to choose our Wannier
functions to be real.
Even if there is no formal proof that at the global minimum of $\Omega$ 
the Wannier functions are real, this is nevertheless a very reasonable 
assumption~\cite{mv},
%
%
%
%
and allows one to fully take advantage of the time-reversal 
symmetry, with significant gain in computational efficiency.

For the minimization of $\Omega$ with respect to the $U^{(\mathbf{k})}_{mn}$
degrees of freedom many efficient schemes are available~\cite{berghold}. 
We decided in this work to implement a damped dynamics algorithm, which allows 
for good control over the process, at the expense of requiring more 
human input for the optimal tuning of the two independent parameters 
(time step and friction).
In antiferromagnetic MnO, a case that is known~\cite{posternak} to be 
difficult to converge, we were able to obtain this way a very accurate and 
symmetrical minimum (to machine precision) in a couple of thousand time 
steps, which required only a few minutes on a modern workstation.
An even more appealing feature of the dynamical scheme is the availability of
a mathematically conserved constant of motion, which provides a very stringent 
test on the accuracy of the implementation.

\section{Convergence properties}

Since the Wannier functions in an insulator are known to be \emph{exponentially}
localized in space~\cite{vanderbiltprl}, similar convergence properties can be expected for any 
physical quantity that is extracted from this particular representation of
the electronic structure.
Instead, as we pointed out at the beginning, both the sum of Wannier centers and
the Berry phase (which is formally related to the Wannier centers by the 
derivation in KSV) converge only as $\mathcal{O}(L^{-2})$, and need special
treatment whenever accurate values are needed.

We will show in this section that this slow convergence is indeed not a
\emph{intrinsic} feature of the ground state electronic structure of an 
extended system, and can be dramatically improved by a simple, 
inexpensive and very general procedure.
Before explaining our correction in detail, we will first provide an intuitive
picture of the position operator in PBC, which, as Resta showed~\cite{restaprl}, 
is the ``kernel'' of both Berry phase and maximally localized Wannier 
function calculations.

Let's consider a one-dimensional system
of one single electronic state $|\psi \rangle$, which we will assume to be 
\emph{well localized} within a periodic cell of length $L$.
The expression for the fundamental, dimensionless complex number $z$ 
is very similar to the 3D expression:
\begin{equation}
z = \langle \psi | e^{i\frac{2\pi}{L}x} |\psi \rangle = |z| e^{i\phi}.
\label{z1d}
\end{equation}
The average value of the
position operator (Eq.~\ref{eqwancen}) becomes:
\begin{equation}
\bar{x} = \frac{L}{2\pi} \mathrm{Im} \ln z = \frac{L}{2\pi} \phi,
\label{barx1d}
\end{equation}
while the quadratic spread (Eq.~\ref{eqspread}) reduces to:
$$
\Omega = \Big( \frac{L}{2\pi} \Big)^2 2( 1-|z|).
$$
%
%
By defining the charge density $\rho(x)=|\psi(x)|^2$, it is
easy to see that the following is true~\footnote{
  The generalization of Eq.~\ref{eqsin} and Eq.~\ref{localpot} to a general 3D
  lattice
  is straightforward by using the set of $\mathbf{b}_i$ vectors and weights
  defined in the text.}:
\begin{subequations}
\begin{equation}
\int_0^L \rho(x) \sin[ \frac{2\pi}{L} (x-\bar{x})] dx = 0
\label{eqsin}
\end{equation}
\begin{equation}
\Omega = \Big(\frac{L}{2\pi} \Big)^2  
\int_0^L \rho(x) \big\{ 2-2\cos [\frac{2\pi}{L} (x-\bar{x}) ] \big\} dx
\label{localpot}
\end{equation}
\end{subequations}
\begin{figure}
\begin{center}
\includegraphics[width=0.4\textwidth]{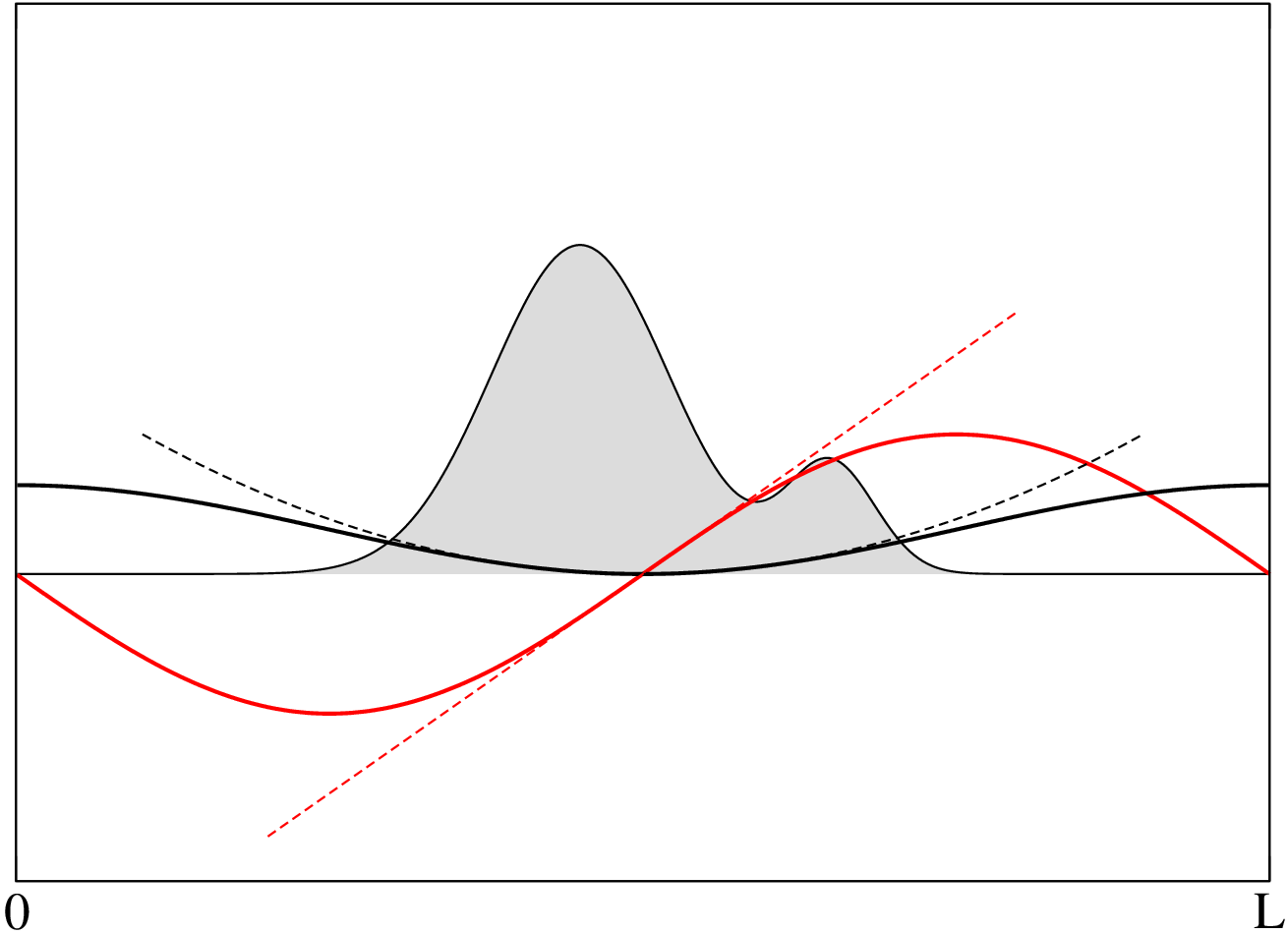}
\end{center}
\caption{Pictorial representation of the position (red)
and spread (black) operators as they are approximated by the Berry-phase 
formalism when working in periodic poundary conditions.}
\label{fig1} 
\end{figure}
These equations can be directly compared to the elementary textbook definitions
of the position and quadratic spread for a square-integrable electronic state in
one dimension (i.e. without PBC, the superscript $F$ stays for ``free-space''):
\begin{subequations}
\begin{equation}
\int_{-\infty}^{\infty} \rho(x) (x-\bar{x}^F) dx = 0
\end{equation}
\begin{equation}
\Omega^F =   
\int_{-\infty}^{\infty} \rho(x) (x-\bar{x}^F)^2 dx
\end{equation}
\end{subequations}
The resemblance is indeed striking, the only difference being the replacement 
of the $x$ and $x^2$ operators with trigonometric functions that are periodic on
the cell.
This relationship between the (polynomial) free-space operators and the
(trigonometric) PBC ones is made evident in Fig.~\ref{fig1}, where they are plotted 
together in order to show their close matching in a region surrounding
the localized state.
Indeed, by a Taylor expansion one obtains:
$$
\Big(\frac{L}{2\pi} \Big) \sin(\frac{2\pi}{L}x) \sim
x - \Big(\frac{2\pi}{L}\Big)^2 \frac{x^3}{3!} + ...
$$
$$
\Big(\frac{L}{2\pi} \Big)^2 \big[ 2-2\cos (\frac{2\pi}{L}x) \big] \sim
x^2 - 2\Big(\frac{2\pi}{L}\Big)^2 \frac{x^4}{4!} + ...
$$
Thus, we arrived from a different starting point at the same 
$\mathcal{O}(L^{-2})$ convergence of the position and spread, which has 
already been discussed in the literature~\cite{mv,puma}.
It is particularly clear from this derivation that the intrinsic property
of the periodic lattice and the localized state are by no means responsible 
for the slow convergence, which is instead determined exclusively by the 
mathematical form of the PBC position operator.

To end this section,
we note that Eq.~\ref{eqsin} alone is not sufficient to \emph{define} 
the center $\bar{x}$, since also $\bar{x}+\frac{L}{2}$ satisfies the same
requirement. 
In the context of Equations~\ref{eqsin} and~\ref{localpot}, a correct
definition of $\bar{x}$ can be given as the points in the lattice which
minimize Eq.~\ref{localpot} (it is easy to show that this
definition is identical to the standard one in Eq.~\ref{barx1d}).
Interestingly, from this point of view the position $\bar{x}$ can be thought
of as an \emph{internal} parameter of the formalism, which is implicit in the 
definition of the spread.

\section{Correction scheme}

Since we are working with Wannier functions which are expected to be well
localized in space (as the 1D state depicted in Fig.~\ref{fig1}), there is actually
no need to insist on using the PBC formulas for calculating Wannier centers.
One could argue here that our ``Wannier functions'' are formally still periodic
(although represented on a large BvK supercell), and since their Hilbert space is 
defined within PBC, only the action of PBC-allowed operators is justified on 
them.
Actually, another point of view can be used. We recall that \emph{true} Wannier
functions are continuous functions in the full 3D (reciprocal) $q$-space.
The mean value of a local operator $V(\mathbf{r})$ in real space can be written
\begin{equation}
\langle w_n | V | w_n \rangle = \int_{All-space} |w_n(\mathbf{r})|^2 
                                V(\mathbf{r}) d \mathbf{r}
\label{eqrspace}
\end{equation}
or, equivalently in $q$-space ($\tilde{\rho}_n(\mathbf{q})$ is the continuous Fourier transform
of the Wannier density $\rho_n(\mathbf{r})=|w_n(\mathbf{r})|^2$):
\begin{equation}
\langle w_n | V | w_n \rangle = \int_{|\mathbf{q}|<E_{cut}} 
                \tilde{V}(\mathbf{q})^*  \tilde{\rho}_n(\mathbf{q}) 
                   d \mathbf{q}
\label{eqqspace}
\end{equation}
If the Wannier function is localized (i.e. zero beyond a given distance from its center), the 
integral in Eq.~\ref{eqrspace} can be limited to a finite region of space, for example a 
cubic box centered around the region where the Wannier density is nonzero.
The $q$-space integral in Eq.~\ref{eqqspace} can then be recast to a sum over a \emph{discrete} set of 
reciprocal space vectors, which is also \emph{finite} because of the plane-wave cut off, and the
result is still \emph{exact}.

If the integration box is chosen to be smaller than the region where the 
Wannier density is nonzero, then the reciprocal-space sum carries an error which is due to 
the overlap between the tails of the Wannier functions and their (artificially) repeated 
images.
This overlap depends on the decay properties of the localized state, and in particular
it goes \emph{exponentially} to zero for increasing integration box size whenever 
$| w_n \rangle$ is exponentially localized.

In a standard DFT simulation of a periodic crystal, the discrete set of reciprocal-space 
Wannier function coefficients are defined by Eq.~\ref{wfgspace}, and converge to their 
thermodynamic limit as soon as the total charge density is converged.
Then, the only effect of a further refinement of the $k$-points mesh is an increase in the
BvK cell volume, which leads to the progressive reduction of the overlap
term discussed above.
Therefore, assuming exponential decay for the Wannier functions,
our technique allows for an \emph{exponential} convergence of 
the calculated expectation value of any free-space operator.
The natural ``bounding box'' for the integration domain in real space is, 
for a general lattice, a Wigner-Seitz BvK cell aligned on the Wannier center.
With this choice, the discrete Fourier representation of a given local free-space operator 
(we use here again the standard conventions for normalizations and Fourier transforms) is:
$$
\tilde{V}(\mathbf{b}) = \frac{1}{V_{BvK}} \int_{Wigner-Seitz} e^{-i\mathbf{b.r}} V(\mathbf{r}) 
            d\mathbf{r},
$$	    
and the expectation value is simply given as
$$
\langle w_n | V | w_n \rangle = V_{BvK} \sum_\mathbf{b} 
  \tilde{V}^*(\mathbf{b}) \tilde{\rho}_n(\mathbf{b}) 
$$

Starting from a well-localized set of Wannier functions we can now 
define a ``refined'' spread operator $\Omega'= \sum_n \Omega_n'$, 
where the contribution from the individual WF is:
$$
\Omega_n' = \int_{Wigner-Seitz} |\mathbf{r} - \bar{\mathbf{r}}_n'|^2 \rho_n(\mathbf{r})
d\mathbf{r}.
$$
The $b$-space expression for this formula can be derived starting
from the Fourier series of a parabola in a one dimensional box:
$$
\frac{1}{L} \int_{-\frac{L}{2}}^{\frac{L}{2}} \cos (\frac{2\pi k}{L}x) x^2 dx =
\Big(\frac{L}{2\pi} \Big)^2 \, \frac{2(-1)^k}{k^2} \qquad (k>0)
$$ 
$$
\frac{1}{L} \int_{-\frac{L}{2}}^{\frac{L}{2}} x^2 dx = \frac{L^2}{4},
$$
and is readily generalized to three dimensions using 
the same set of vectors and weights \{$\mathbf{b}_i,\bar{\mathrm{w}}_i$\} 
introduced in Section 2: 
$$
\Omega_n' = 2 V_{BvK} \sum_{i,k>0} \bar{\mathrm{w}}_i 
\mathrm{Re} \Big[ \frac{2(-1)^k}{k^2} \,
                  e^{i k \mathbf{b}_i . \bar{\mathbf{r}}_n'} \, 
		\tilde{\rho}_n(k\mathbf{b}_i) \Big] +
$$
\begin{equation}				 
	+  \frac{\pi^2}{3} \sum_i \bar{\mathrm{w}}_i
\label{omeganprim}
\end{equation}
The ``refined'' position $\bar{\mathbf{r}}_n'$ which appears in 
Eq.~\ref{omeganprim} is then again an \emph{internal parameter}, 
which is defined by the minimum of $\Omega_n'$ for a given $\tilde{\rho}_n$ 
(see the discussion at the end of Section 3).
By taking the gradient of $\Omega_n'$ with respect to $\bar{\mathbf{r}}_n'$
one obtains that, at the minimum, the integral:
$$
\Delta \mathbf{r}_n = \int_{Wigner-Seitz} (\mathbf{r} - \bar{\mathbf{r}}_n') \rho_n(\mathbf{r})
d\mathbf{r} 
$$
vanishes. 
Consistently with the definition of the spread,
this condition has to be enforced in reciprocal space, where this integral 
becomes:
\begin{equation}
\Delta \mathbf{r}_n = -2 V_{BvK}
\sum_{i,k > 0} \bar{\mathrm{w}}_i \mathbf{b}_i \mathrm{Re} \, \Big[ i \frac{(-1)^k}{k} 
                                 e^{i k \mathbf{b}_i . \bar{\mathbf{r}}_n'} 
				 \tilde{\rho}_n(k\mathbf{b}_i) \Big],
\label{deltar}
\end{equation}
The stationary point can be obtained iteratively starting from a set of 
maximally localized Wannier functions and unrefined centers, 
by updating at every iteration $\bar{\mathbf{r}}_n'$ through the addition of
$\Delta \mathbf{r}_n$ as calculated in Eq.~\ref{deltar} until convergence
is reached.
If the Wannier function is exactly zero in a region surrounding the 
boundary of the Wigner-Seitz cell, 
one iteration is sufficient to provide the \emph{exact} value of the center, 
while for less converged cases up to ten iterations may be necessary to 
achieve machine precision.
These iterations have anyway negligible cost, since the Fourier transform 
of the Wannier function on the
BvK cell has to be evaluated only at the beginning 
(twice for each Wannier function to get the density in reciprocal space).


Both expressions $\Omega$ and $\Omega'$ are in fact particular cases of a class of 
localization criteria which rely on individual Wannier densities only, 
through some generalized spread functional $S$:
$$
\Omega = \sum_n S[\rho_n] 
$$
%
%
A similar generalised, density-dependent spread can be used in
practice to explore
alternative localization criteria, like e.g. the maximal Coulomb self-repulsion of Edmiston and
Ruedenberg~\cite{edmiston}, or the orbital self-interaction as defined by Perdew and 
Zunger~\cite{perdew}. 
An article comparing such alternatives is under preparation~\cite{inprep}.

Since the present free-space-like expressions for position and spread are more 
accurate than those derived in the first part of this work, 
one could wonder why we did not use them from the beginning.
The reason is exclusively related to computational efficiency. In Eqns.~\ref{eqnsk} 
the localization algorithm involves operations on small $J \times J$ matrices only,
where $J$ is the number of bands in the primitive cell (the computationally intensive
calculation of the $M^{(\mathbf{k}, \mathbf{b}_i)}$ matrices has to be performed
only once at the beginning of the iterative minimization).
If, instead, the refined spread (or one of the alternative localization criteria 
discussed above) is used directly for localizing the Wannier functions, several Fourier 
transforms on the full Wannier (BvK) grid are required for each iteration, at a substantially 
higher cost.
This expensive procedure is anyway not necessary for the scope 
of the present work, since the actual set of Wannier wavefunctions 
obtained from one localization method or the other coincide in practice
to a high degree of accuracy~\cite{hydrogen} (in particular, 
the decay properties are expected to be very similar).
Therefore, we find it most convenient to use this refinement step in a one-shot fashion once a set
of maximally localized Wannier functions is obtained within the more efficient localization 
functional (Eqns.~\ref{eqnsk}).

\section{Numerical tests}

\begin{figure}
\begin{center}
\includegraphics[width=0.4\textwidth]{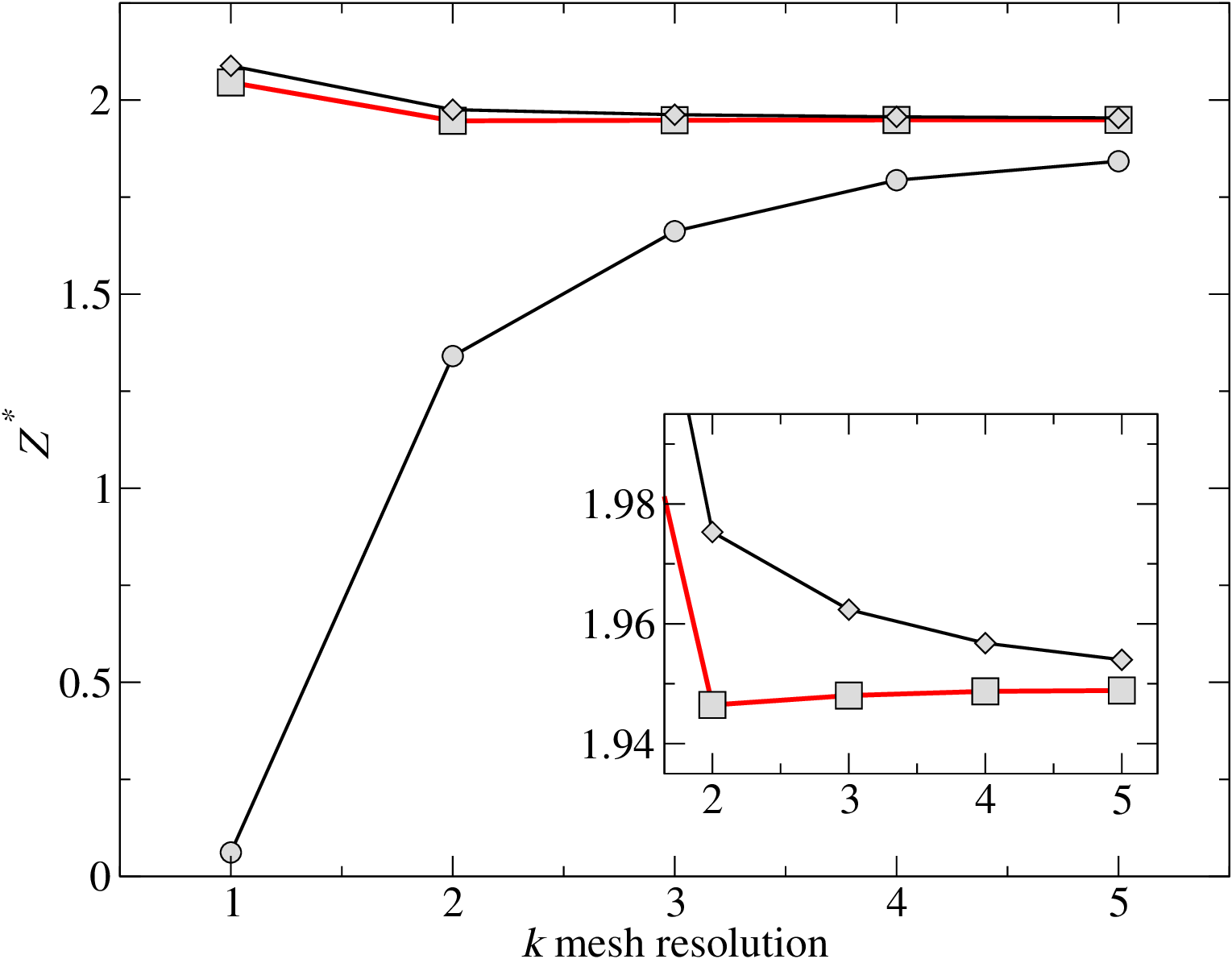}
\end{center}
\caption{Convergence of the Born effective charge $Z^*$ of oxygen in MgO
as computed using the sum of unrefined Wannier centers (Eq.~\ref{wancenk}, circles),
the Berry-phase approach (Eq.~\ref{berry}, diamonds) and our new scheme 
(Eq.~\ref{omeganprim}, squares). 
The blow-up in the inset shows the improvement of our method
with respect to the Berry-phase
approach.}
\label{figborn}
\end{figure}


To demonstrate the effectiveness of our method we have chosen two examples 
which have been extensively studied in the literature: (i) the dynamical Born effective 
charge
of oxygen atoms in rocksalt MgO, and (ii) the spontaneous polarization
of the ferroelectric, tetragonally distorted phase of KNbO$_3$.
Our calculations were performed within the local density approximation~\cite{perdew},
by using norm-conserving Troullier and
Martins~\cite{tm} pseudopotentials in the Kleinman and Bylander~\cite{kb}
form. A non-linear core correction~\cite{louie} was adopted for the Mg pseudopotential, 
while the K pseudopotential was generated in the $4s^0$ ionized configuration
with the semicore $3s,3p$ orbitals included in the valence. 
We used the experimental lattice constants and atomic positions ($a_0$=7.96 a.u. for 
MgO~\cite{mgoexp}, 
and the structural data for KNbO$_3$ from Ref.~\onlinecite{baldoknbo3}). 
We expanded the electronic ground state on a plane-wave basis up to a cutoff of 70 Ry.
The BZ sampling was performed with $\Gamma$-centered simple cubic (orthorombic) grids in 
reciprocal space for MgO (KNbO$_3$), by taking into account the time-reversal symmetry only.

We will compare the results as a function of $k$-mesh resolution for three different methods
for calculating the polarization:
(i) the sum of Wannier centers as obtained by Eq.~\ref{wancenk};
(ii) the sum of \emph{refined} Wannier centers as described in the previous
section; (iii) the Berry-phase approach.
We note that the Berry-phase result can be readily obtained from the quantities that are
already available in the localization formalism:
\begin{equation}
\mathbf{r}_{Berry} = 
- \frac{1}{N} \sum_i \bar{\mathrm{w}}_i \mathbf{b}_i \sum_\mathbf{k} \mathrm{Im} \ln 
\det M^{(\mathbf{k,b}_i)}.
\label{berry}
\end{equation}

\subsection{MgO}

\begin{figure}
\begin{center}
\includegraphics[width=0.4\textwidth]{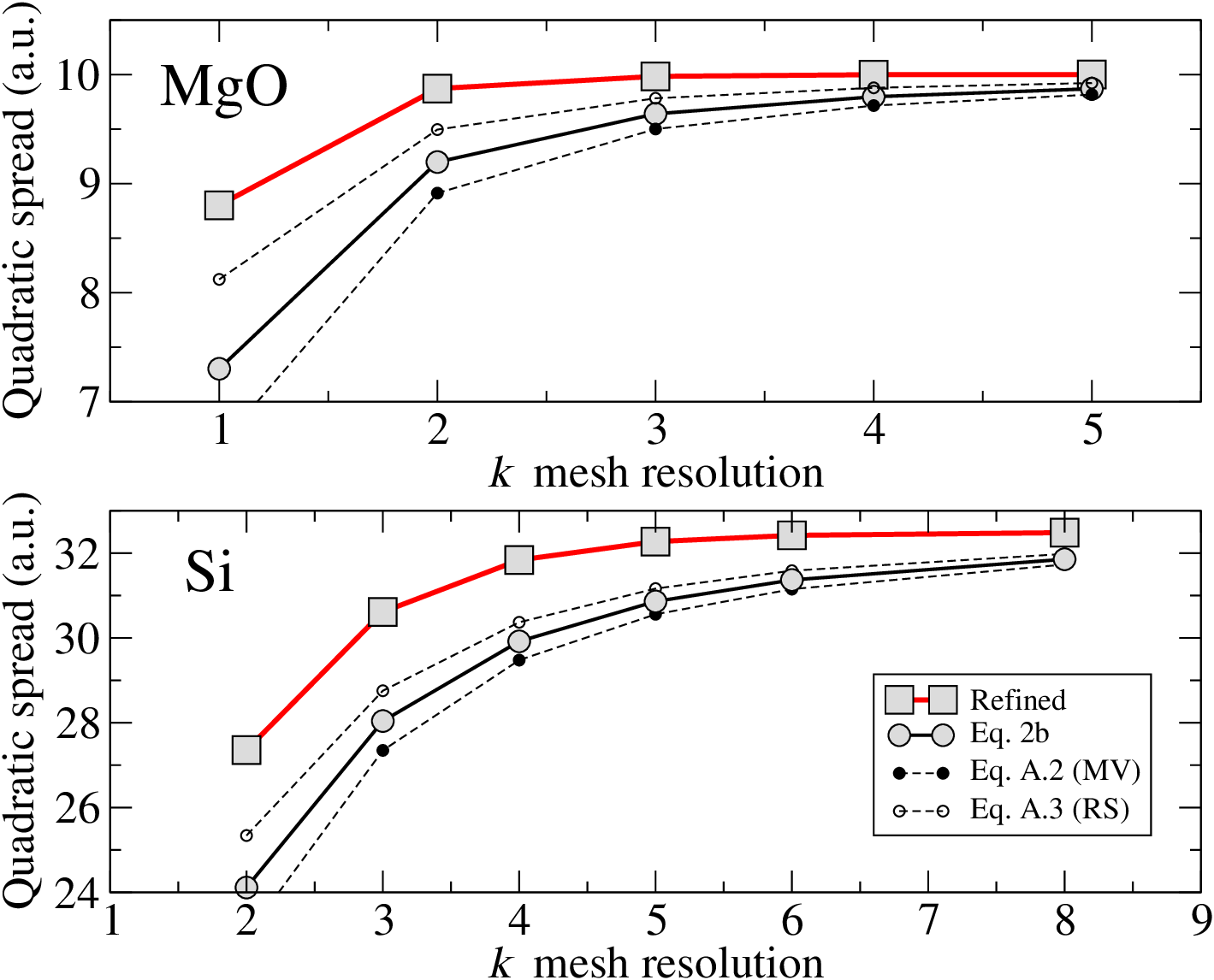}
\end{center}
\caption{Convergence of the total spread in the case of MgO and Si. 
The improvement provided by the refined value 
(Eq.~\ref{omeganprim}, red line and squares) with respect to the
standard ``trigonometric'' spread (Eq.~\ref{spreadk}, black line and circles)
is apparent.
We provide for comparison the alternative spreads discussed in the Appendix,
which are also based on the same trigonometric kernel and show
similar, slow convergence properties.}
\label{figspread}
\end{figure}

\begin{figure}[b!]
\begin{center}
\includegraphics[width=0.45\textwidth]{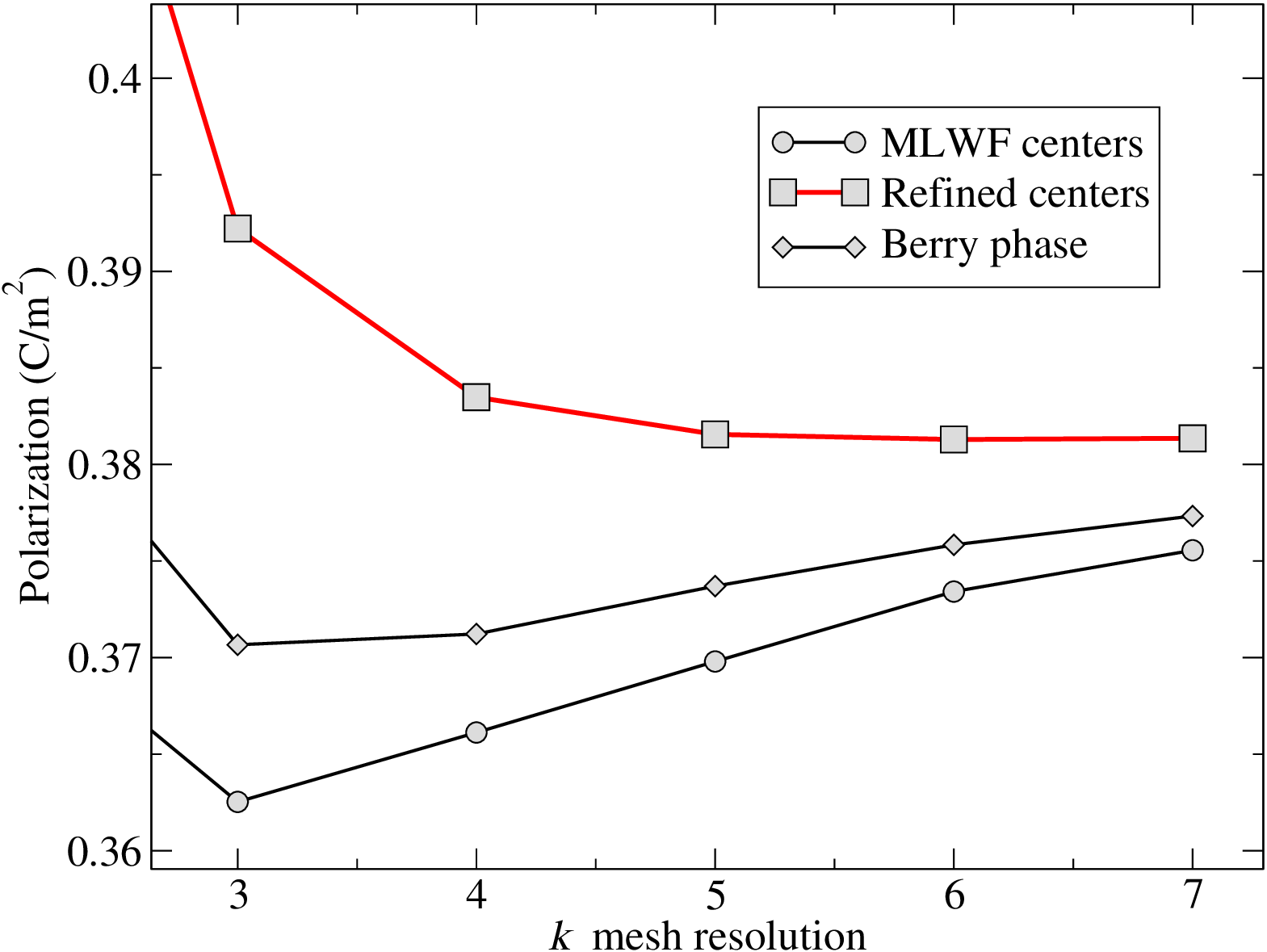}
\end{center}
\caption{Convergence of the spontaneous polarization of the (tetragonal)
ferroelectric phase of KNbO$_3$,
as computed using the sum of unrefined Wannier centers (Eq.~\ref{wancenk}, circles),
the Berry-phase approach (Eq.~\ref{berry}, diamonds) and our new scheme 
(Eq.~\ref{omeganprim}, squares). 
}
\label{figknbo3}
\end{figure}

The dynamical Born effective charge ($Z^*$) of oxygen was calculated by the 
finite-difference method,
i.e. by considering the difference in total polarization between the ideal centrosymmetric
ground state and an atomic configuration where the oxygen sublattice was displaced 
by 1\% of the cubic lattice constant along the $x$ direction.
The atomic coordinates were prepared in such a way that, in the ideal lattice, the O atom
sits at the origin, and in this case the electronic contribution to the polarization is
exactly zero modulo a polarization quantum (all four Wannier functions are symmetrical 
about the O in this case).
%
%
We compare in Fig.~\ref{figborn} the
resulting value for $Z^*$ as calculated
by the three different methods (i-iii).
The results clearly show that the sum of the 
\emph{unrefined} Wannier centers can be very inaccurate, and even for the 
finest mesh the error is still large. 
MgO is probably a very unfortunate case in that each $sp^3$-like Wannier function
has a strongly asymmetric shape, and the errors in the individual centers do not 
cancel out efficiently in the strained configuration, so that the total polarization 
carries an important deviation from the exact value.
The Berry-phase calculation is a much better estimate, but in the inset 
it can be seen that the convergence is still relatively slow.
As we explained in the preceding sections, it was already shown that the Berry-phase result 
converges only as $\mathcal{O}(L^{-2})$, i.e. it shares the same asymptotic behaviour as the sum
of the unrefined Wannier centers (\emph{albeit} with a quite different
prefactor in this particular case).
The sum of the \emph{refined} Wannier centers instead shows an extremely 
fast convergence, and gives a very accurate result already for a
$2 \times 2 \times 2$ mesh.
The value of $Z^*$ we obtain is -1.95, which is in excellent agreement with previous 
experimental and theoretical investigations~\cite{pumaprl}.

To complement our methodological test, we calculated also the refined value for the 
total quadratic spread as a function of $k$-mesh resolution, and the results are reported
in Fig.~\ref{figspread}.
It is clear that this quantity shares the same, excellent, convergence properties as the 
position operator (upper panel).
In the lower panel of the same figure we report for comparison the results of an analogous 
calculation of the total spread in bulk silicon.
The convergence is slower than in MgO, as can be expected from the very different character
of this covalent compound as compared to the highly ionic magnesium oxide, but the benefit 
that can be obtained through the use of the more accurate free-space definition of the 
spread is still very clear.
The ``unrefined'' value of the spread is also compared to the alternative,
very similar prescriptions discussed in the Appendix. 

\subsection{KNbO$_3$}

We present in Fig.~\ref{figknbo3} our results for the spontaneous polarization of KNbO$_3$.
The sum of the unrefined Wannier centers is less inaccurate in this case,
and is fairly close to the values obtained within the Berry phase formalism.
The sum of the refined centers has, again, much better convergence properties than the
two traditional methods. 
By increasing 
the mesh from $6 \times 6 \times 6$ to $7 \times 7 \times 7$ the value of the spontaneous
polarization increases by 0.02 \%, while within the same $7 \times 7 \times 7$ mesh the 
traditional techniques carry an error which is two orders of magnitude higher.
Extrapolating the $\mathcal{O}(L^{-2})$ trend one can guess that $\sim 70$ $k$-points along
the reciprocal space stripes~\cite{kingsmith} would be needed to achieve similar accuracy 
within the Berry-phase formalism.
The final value we obtain, 0.38 C/m$^2$, compares very well with experimental data and 
previous theoretical 
investigations~\cite{baldoknbo3,dovesiknbo3,restaknbo3,vanknbo3}.

\section{Conclusions}

In conclusion, we have derived a simple and general formalism for the
computation of maximally localized Wannier functions. 
We provide an intuitive picture of the convergence properties of this scheme
and similar ones, relating them to the
Taylor expansion of elementary trigonometric functions.
We show that the convergence can be dramatically improved by a
simple strategy based on the exponential localization of the Wannier functions
in insulating materials.
%
%
%
%
%
We expect our scheme to open the way to both accurate and efficient calculations
of polarization properties in a wide range of physical systems, making the 
expensive linear-response approach or the relatively cumbersome 
non-self consistent calculation of ``stripes'' in reciprocal space 
unnecessary.  

As a final remark, we note that the Wannier function-based theory of 
polarization is becoming
increasingly important especially in disordered systems, where not only
the global polarization but also the \emph{local} bonding properties and dipole 
moments may be interesting to follow during, 
e.g. a molecular dynamics simulation~\cite{sharma}.
In these applications the improved accuracy provided by our method could
be an extremely valuable tool.

We wish to thank David Vanderbilt for insightful comments on the manuscript.
This work was supported by the National Science Foundation's Division of
Materials Research through the Information Technology Research program,
grant number DMR-0312407, and made use of MRL Central Facilities
supported by the National Science Foundation under award No. DMR-0080034

\appendix*

\section{Decomposition into invariant, off-diagonal and diagonal parts}

The form~\ref{eqspread} for the spread functional was chosen
mainly because of its simplicity, and because it allows for
a direct interpretation as the integral of the Wannier density multiplied by
a real function on the BvK cell (see the discussion in Sec.~3).
Unfortunately this expression does not lead 
to an elegant separation into invariant and non-invariant parts. 
However, this issue is readily solved by choosing an alternative definition
of the spread:
%
\begin{equation}
\Omega_{MV} = \Big(\frac{L}{2\pi} \Big)^2 (1-|z|^2),
\label{spreadmv}
\end{equation}
which coincides with the $\Gamma$-point prescription of MV and which
\emph{does} allow for an exact separation of the invariant part.
This choice still allows for the simple interpretation based on cosine-like
functions. If we define a function of $x_0$:
\begin{equation}
f(x_0) =  
\int_0^L \rho(x) \cos [\frac{2\pi}{L} (x-x_0) ]  dx
\end{equation}
it is clear that when 
$x_0$ maximizes $f$, $x_0$ is automatically the Wannier center of
Eq.~\ref{eqwancen}. Both expressions for the spread ($\Omega$ as in
Eq.~\ref{eqspread} and $\Omega_{MV}$ discussed here) are consistent with
the same value of $x_0$ at the minimum:
$$\Omega =\Big(\frac{L}{2\pi} \Big)^2 \min_{x_0} 2[1-f(x_0)]$$ 
$$\Omega_{MV} =\Big(\frac{L}{2\pi} \Big)^2 \min_{x_0}  [1-f^2(x_0)]$$ 

%
Moving on to 3D, the operational definition of the spread becomes:
$$
\Omega_{MV}  = \sum_n \sum_i \bar{\mathrm{w}}_i \, 
\big(1 - | z_n^{(i)}|^2 \big),
$$
where it is easy to see that $z_n^{(i)}$ are nothing other
than the matrix elements indicated as $X_{nn}$, $Y_{nn}$, $Z_{nn}$ in MV.

Now, ``folding'' this expression in $k$-space leads to a formula which is 
similar to Eq.~\ref{spreadk}:
$$
\Omega  = \sum_n \sum_i \bar{\mathrm{w}}_i  \, \big(1 - \Big| 
          \frac{1}{N} \sum_\mathbf{k} M^{(\mathbf{k,b}_i)}_{nn}
          \Big|^2 \big)
$$

Thinking in terms of the big BvK cell, this can be written equivalently
as:
$$
\Omega  = \frac{1}{N} \sum_i \bar{\mathrm{w}}_i (NJ - \sum_{\mathbf{R},n}
|\langle \mathbf{R} n| e^{-i\mathbf{b}_i.\mathbf{r}} 
| \mathbf{R} n \rangle |^2),
$$
where the leading factor $1/N$ gives the spread per \emph{primitive} cell,
and the same notations as MV for the $n$-th Wannier function 
at the $\mathbf{R}$ site, $| \mathbf{R} n \rangle $ are used.
From this expression it is clear how to construct an obvious invariant
quantity, $\Omega_I$ ($J$ is the number of bands in the primitive cell):
$$
\Omega_I  = \frac{1}{N} \sum_i \bar{\mathrm{w}}_i (NJ - 
\sum_{\mathbf{R}\mathbf{R}',nm}
|\langle \mathbf{R} n| e^{-i\mathbf{b}_i.\mathbf{r}} 
| \mathbf{R}' m \rangle |^2),
$$
and what remains to do is to ``unfold'' this
formula in $k$-space.
A first simplification is trivial:
$$
\Omega_I  = \sum_i \bar{\mathrm{w}}_i (J - 
\sum_{\mathbf{R},nm}
|\langle \mathbf{R} n| e^{-i\mathbf{b}_i.\mathbf{r}} 
| \mathbf{0} m \rangle |^2).
$$
A second simplification is obtained by reversing the formula between Eq. 5
and 6 of MV, leading to:
$$
\Omega_I  = \sum_i \bar{\mathrm{w}}_i (J - 
\sum_{\mathbf{R},nm}
\Big|\frac{1}{N} \sum_\mathbf{k} e^{i \mathbf{k.R}} 
\langle u_{n\mathbf{k}}| e^{-i\mathbf{b}_i.\mathbf{r}} 
| u_{m\mathbf{k+b}} \rangle \Big|^2).
$$
By writing explicitly $|z|^2 = z^*z$ and noticing that 
$$ \sum_\mathbf{R} e^{i \mathbf{(k-k').R}} = N \delta_\mathbf{k,k'}, $$
we obtain the final expression in $k$-space:
\begin{equation}
\Omega_I  = \sum_i \bar{\mathrm{w}}_i (J - \sum_{mn} \frac{1}{N} \sum_\mathbf{k} 
\Big|M^{(\mathbf{k,b}_i)}_{nm} \Big|^2),
\end{equation}
which is exactly Eq. 34 of the MV paper.

It is interesting to work out the remaining terms, $\Omega_D$ and 
$\Omega_{OD}$, which are indicated as ``diagonal'' and ``off-diagonal''
parts in MV (we recall that $\Omega_D$ vanishes for a centrosymmetric
system~\cite{mv}). The easiest way is to first solve the expression for
$$
\Omega_{MV} - \Omega_D = \sum_i \bar{\mathrm{w}}_i (J - 
\sum_{\mathbf{R},n}
|\langle \mathbf{R} n| e^{-i\mathbf{b}_i.\mathbf{r}} 
| \mathbf{0} n \rangle |^2).
$$
By using an analogous algebra we readily arrive at the formula in $k$-space:
$$
\Omega_{MV} - \Omega_D = \sum_i \bar{\mathrm{w}}_i (J - \sum_{n} \frac{1}{N} 
\sum_\mathbf{k} 
\Big|M^{(\mathbf{k,b}_i)}_{nn} \Big|^2),
$$
from which it is very easy to evaluate $\Omega_{OD}$:
$$
\Omega_{OD} = \sum_i \bar{\mathrm{w}}_i \sum_{m \neq n} \frac{1}{N} 
\sum_\mathbf{k} 
\Big|M^{(\mathbf{k,b}_i)}_{mn} \Big|^2.
$$
This means that $\Omega_D$ is given by the following difference:
$$
\Omega_D = \sum_{i,n} \bar{\mathrm{w}}_i \big( \Big| 
          \frac{1}{N} \sum_\mathbf{k} M^{(\mathbf{k,b}_i)}_{nn}
          \Big|^2 -
	   \frac{1}{N} 
\sum_\mathbf{k} 
\Big|M^{(\mathbf{k,b}_i)}_{nn} \Big|^2 \big).
$$  

Comparing this formalism with the MV one, it is clear that $\Omega_{I}$ and
$\Omega_{OD}$ are identical, while the terms $\Omega$ and $\Omega_D$ differ.
This derivation provide in a certain sense a ``unification'' of the
two, formerly distinct, MV prescriptions for the $\Gamma$-point case and 
in $k$-space.
The gradients with respect to unitary rotations of the Bloch orbitals 
are simply given by setting $C=z$ (instead of $C=z/|z|$) in Eq.~4.

To complete our discussion, we note that a third form of the one-particle 
quadratic spread was proposed by Resta and Sorella~\cite{rs}, which
leads to yet another operational definition for the localization 
criterion:
$$
\Omega_{RS}  = -\sum_n \sum_i \bar{\mathrm{w}}_i \, 
\ln \, |z_n^{(i)}|^2.
$$
The $k$-space folding of this formula is straightforward, while the gradient
is again given by Eq.~4, with $C=z/|z|^2$.
All functionals $\Omega$, $\Omega_{MV}$ and $\Omega_{RS}$ are identical in
the thermodynamic limit. For finite BvK cells they all share the same 
definition of the Wannier center. 
The resulting maximally localized Wannier functions themselves are identical 
in cases where $|z_n|$ are equal for all $n=1,...,J$ bands (e.g. bulk Si, 
centrosymmetric MgO crystal). 
The numerical value of the spread can differ slightly, because the higher
orders in the Taylor expansion are different. 
Some examples concerning this discrepancy are reported in the main text (see,
e.g., Fig.~\ref{figspread}).

\end{document}